\documentclass[useAMS,usenatbib]{mn2e}

\usepackage{amsmath,natbib,graphicx}
\usepackage{epsf}
\usepackage{epstopdf}
\input{epsf}
\usepackage{epsfig}
\usepackage{color}
\usepackage{subfigure}

\DeclareGraphicsExtensions{.jpg,.pdf,.png,.eps,.ps}

\newcommand{\HI}{H\:\!\textsc{i}}
\newcommand{\sfrA}{SFR$_{{\rm H}\alpha}$}

\title[Metallicity, gas content, and stellar mass]{A fundamental relation between the metallicity, gas content, and stellar mass of local galaxies}
\author[M.\,S.\ Bothwell et al. ]
{M. S. Bothwell$^{1,2}$\thanks{E-mail:
matthew.bothwell@gmail.com}, 
R. Maiolino$^{1,2}$,
R. Kennicutt, Jr.$^{3}$,
G. Cresci$^{4}$,
F. Mannucci$^{4}$, \newauthor
A. Marconi$^{5}$,
C. Cicone$^{1,2}$
\\ 
$^{1}$Cavendish Laboratory, University of Cambridge, 19 J.J. Thomson Avenue, Cambridge, CB3 0HE, UK\\
$^{2}$Kavli Institute for Cosmology, University of Cambridge, Madingley Road, Cambridge CB3 0HA, UK\\
$^{3}$Institute of Astronomy, University of Cambridge, Madingley Road, Cambridge, CB3 0HA\\
$^{4}$INAF, Osservatorio Astrofisico di Arcetri, Largo E. Fermi 5, 50125 Firenze, Italy\\
$^{5}$Universit\`{a} di Firenze, Dipartimento di Fisica e Astronomia, Via G. Sansone 1, I-50019 Sesto Fiorentino, Firenze, Italy
}
\begin{document}
\date{Accepted ----. Received ---- in original form ----}

\pagerange{\pageref{firstpage}--\pageref{lastpage}} \pubyear{2012}

\maketitle
\begin{abstract}
Recent results have suggested that the well known mass-metallicity relation has a strong dependence on the star formation rate, to the extent that a three dimensional `fundamental metallicity relation' exists which links the three parameters with minimal scatter.  In this work, we use a sample of 4253 local galaxies observed in atomic hydrogen from the ALFALFA survey to demonstrate, for the first time, that a similar fundamental relation (the \HI-FMR) also exists between stellar mass, gas-phase metallicity, and \HI\ mass. This latter relation is likely more fundamental, driving the relation between metallicity, SFR and mass. At intermediate masses, the behaviour of the gas fundamental metallicity relation is very similar to that expressed via the star formation rate. However, we find that the dependence of metallicity on \HI\ content persists to the highest stellar masses, in contrast to the `saturation' of metallicity with SFR. It is interesting to note that the dispersion of the relation is very low at intermediate stellar masses ($\rm 9<\log{(M_*/M_{\odot})}<11$), suggesting that in this range galaxies evolve smoothy, in an equilibrium between gas inflow, outflow and star formation. At high and low stellar masses, the scatter of the relation is significantly higher, suggesting that merging events and/or stochastic accretion and star formation may drive galaxies outside the relation. We also assemble a sample of galaxies observed in CO. However, due to a small sample size, strong selection bias, and the influence of a metallicity-dependent CO/H$_2$ conversion factor, the data are insufficient to test any influence of molecular gas on metallicity. 

\end{abstract}


\section{Introduction}

Understanding the physical processes regulating the metallicity of galaxies is one of the most important problems in modern astrophysics. The gas-phase metallicity functions as a fossil record of a galaxy's history, encapsulating information about both the past star formation and the infall/outflow of interstellar gas.

It has long been known that metallicity correlates well with both bolometric luminosity and stellar mass, with the latter (the `mass-metallicity' relation) in particular being extensively studied in recent years, both locally (e.g. \citealt{1979A&A....80..155L}; \citealt{2004ApJ...613..898T}) and at higher redshifts (e.g. \citealt{2005ApJ...635..260S}; \citealt{2006ApJ...644..813E}; \citealt{2008A&A...488..463M}; \citealt{2009MNRAS.398.1915M}; \citealt{2012MNRAS.421..262C}). A picture has emerged of a well-constrained local relation which evolves strongly out to $z>3$, placing valuable constraints on galaxy evolution models. 

A host of physical models has been proposed to explain the mass metallicity relation, including star formation efficiency issues \citep{2007ApJ...655L..17B} and IMF variations \citep{2007MNRAS.375..673K}. Perhaps unsurprisingly, though, most explanations for the relation have focussed on explanations involving the global behaviour of gas (i.e. \citealt{1976ARA&A..14...43A}), which can serve both as a pristine infalling diluent (e.g. \citealt{Dave:2010aa}; \citealt{2008MNRAS.385.2181F}) and as a metal rich outflow (\citealt{2010A&A...514A..73S}; \citealt{2011MNRAS.417.2962P}). 

More recently, it was found that the mass-metallicity relation can be seen as a consequence of a more fundamental relation between stellar mass, metallicity, and star formation rate (SFR) (\citealt{2010MNRAS.408.2115M}; \citealt{2010A&A...521L..53L}; \citealt{2013ApJ...765..140A}). This `fundamental metallicity relation' (`FMR' hereafter) was found to exhibit remarkably little scatter, and -- more importantly -- was found not to evolve out to at least $z=2.5$ (\citealt{2010MNRAS.408.2115M}; \citealt{2010A&A...521L..53L}). As metallicity anti-correlates with SFR at a constant stellar mass, the apparent evolution of the mass metallicity relation can be explained as a natural consequence of the fact that samples of galaxies at higher redshifts have elevated star formation rates relative to local galaxies. If the FMR is seen as a relation in a 3D parameter space defined by stellar mass, metallicity, and SFR, samples of galaxies at different redshifts simply inhabit different regions of this unchanging `fundamental plane'. 

The strength of the correlation between these three parameters is at first glance somewhat surprising, and strongly suggests the presence of highly efficient regulatory mechanisms. \cite{2010MNRAS.408.2115M} attribute the existence of the FMR to a well-regulated balance between pristine inflows and enriched outflows. In this model, metal-poor gas falls into a galactic halo, and acts to both boost star formation and decrease metallicity by a `dilution' effect. Likewise, star formation-driven outflows remove metal-rich gas at a rate which is strongly dependent on stellar mass. These two effects combine, resulting in a tight correlation between the three parameters stellar mass, metallicity, and SFR. These mechanisms have been further explored by \cite{2012MNRAS.421...98D} and \cite{2012arXiv1202.4770D}, who each present a set of simple equations which link the observable quantities of the FMR to the behaviour of infalling and outflowing gas. The difference between these models is that \cite{2012MNRAS.421...98D} assume an `equilibrium condition', whereby the inflow rate is balanced by the sum of the star formation and outflow rates (this is sometimes referred to as the `reservoir' or `bathtub model' -- see \citealt{2010ApJ...718.1001B}; \citealt{Krumholz:2011aa}). 

The common thread underpinning all physical and analytical models of the FMR is the understanding that the driving force behind these tight correlations is the behaviour of gas. In particular, as mentioned above, the anti-correlation between metallicity and SFR is thought to be a byproduct of a more fundamental relation between metallicity, gas content, and stellar mass. It has already been noted by authors that there is a connection between metallicity and gas content; \cite{1979A&A....80..155L} found an inverse correlation between heavy element abundance and gas fraction for local blue compact galaxies. Both \cite{1996ApJ...462..147S} and \cite{2008ApJ...685..904P} noted that local metal rich galaxies tended to be gas-poor, and \cite{2006PASP..118..517B} made the same observation for cluster galaxies. Recently, \cite{2013A&A...550A.115H} systematically analysed the influence of \HI\ gas fractions on the mass-metallicity relation for a sample of 260 local galaxies, finding that the gas fraction is closely related to the scatter in the  mass-metallicity relation. \cite{2012ApJ...748...48R} also found a connection between gas-phase metallicity and \HI\ deficiency for a small sample of 6 local cluster galaxies, with \HI\ deficient galaxies being more metal rich by $\sim 0.1$ dex. 

In this paper, we use a large sample of local galaxies to reveal the existence of a fundamental relation between stellar mass, metallicity, and atomic gas content. We also explore the relation between stellar mass, metallicity, and molecular gas content; however, such a relation is much more difficult to investigate due to the limited data available, selection effects, and issues associated with the CO/H$_2$ conversion factor. The paper is structured as follows: in \S2 we present the samples used in this analysis, and describe the method used to determine metallicity. \S3 gives our results, and we discuss the implications in \S4. We present our conclusions in \S5.   

\begin{figure*}
\centering
\mbox
{
  \subfigure{\includegraphics[width=8.5cm]{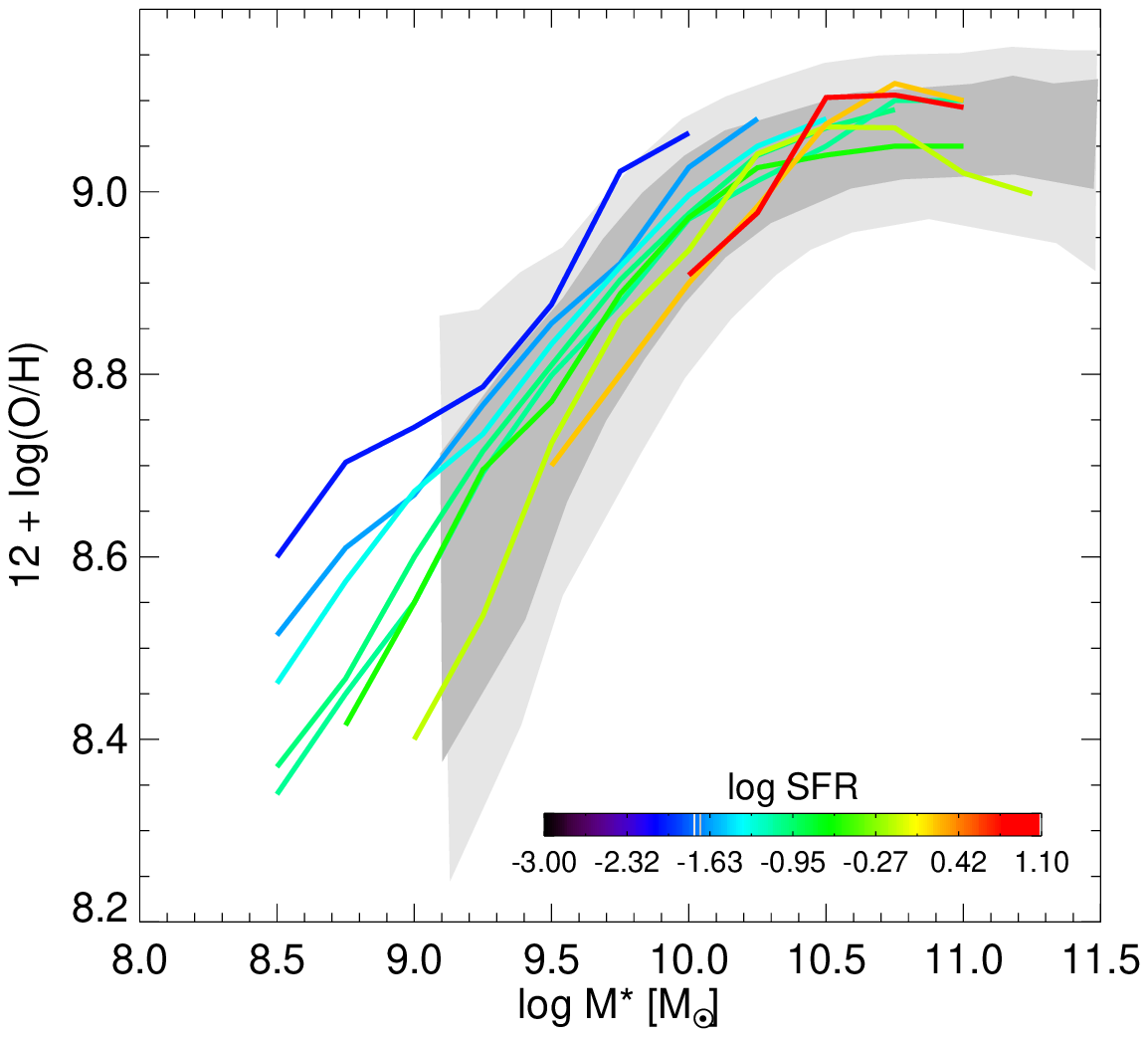}}
  \subfigure{\includegraphics[width=8.5cm]{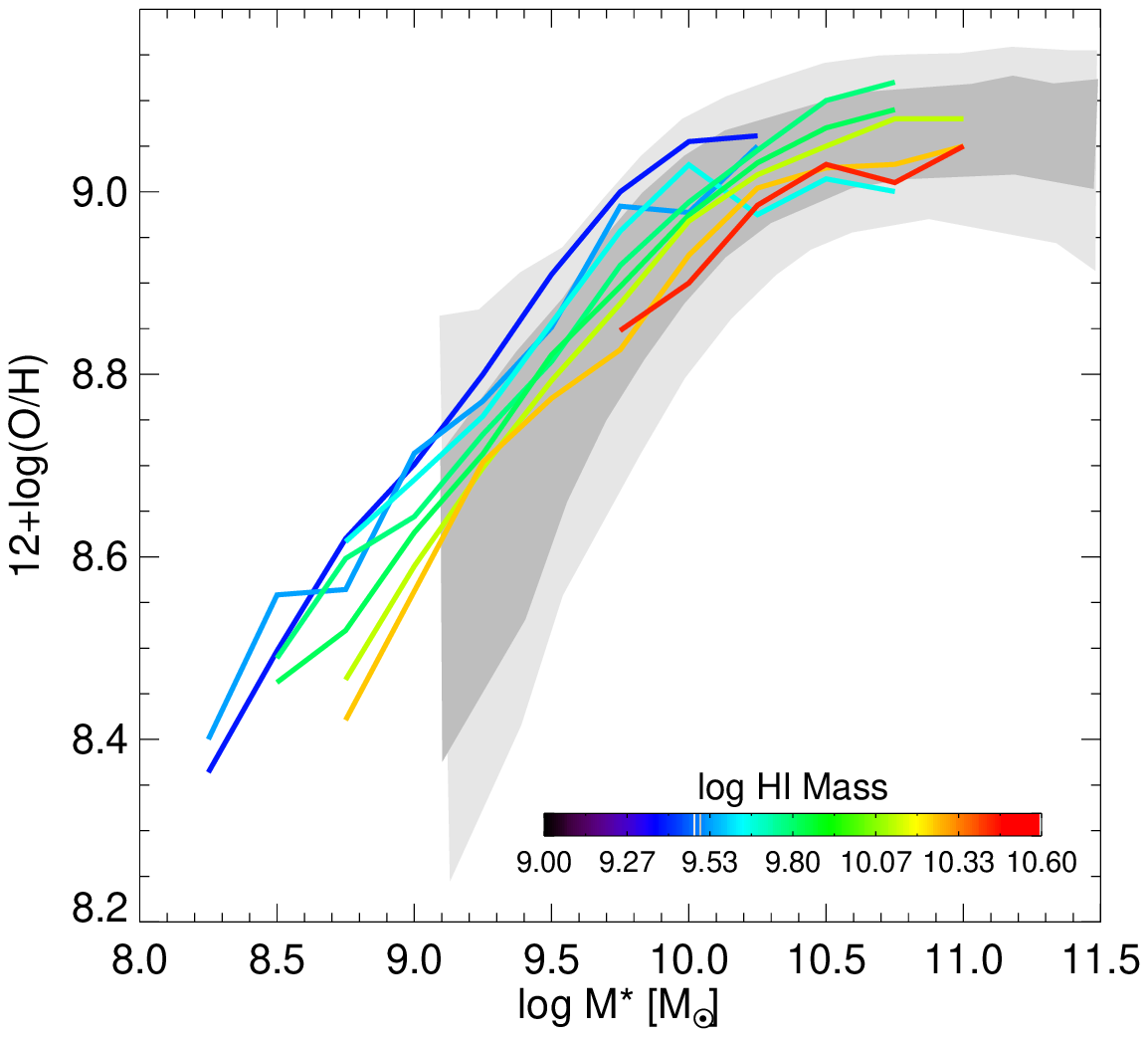}} 
}
\caption{The mass-metallicity relation for the 4253 ALFALFA galaxies in our sample. The grey shaded areas show the area that contains 64\% (light shaded area) and 90\% (dark shaded area) of all SDSS galaxies from the Mannucci et al. (2010) study. The coloured lines show the mean trends for galaxies in bins of H$\alpha$-SFR (left panel) and \HI\ mass (right panel). It can be seen that at a constant stellar mass, metallicity is a decreasing function of both SFR and \HI\ mass.}
\label{fig:MZ-lines}
\end{figure*}

\section{Sample selection}
\label{sec:samples}

The primary motivation behind our sample selection was assembling a large number of galaxies with the full complement of data, sufficient for measuring the gas mass, the stellar mass, the gas-phase metallicity, and the star formation rate in a homogeneous manner across the sample. By far, the most time consuming of these parameters to observationally measure is the gas content. While surveys such as the SDSS now routinely measure the optical properties of millions of $z\sim0$ galaxies, obtaining hydrogen masses for similarly large numbers of galaxies remains more of a challenge. While large, modern surveys for atomic hydrogen can now collect data for (on the order of) $10^4$ galaxies, molecular hydrogen remains more elusive yet, with fewer than $10^3$ galaxies having the CO observations required to extrapolate a molecular gas mass. 

Below, we discuss the surveys we have used to build our samples of local galaxies. 

\subsection{HI sample}
\label{sec:hi}

To construct our \HI\ sample, we used the first data release of the `Arecibo Legacy Fast ALFA' survey (ALFALFA), a wide-field survey of \HI\ in the local Universe (\citealt{2011AJ....142..170H}; see also \citealt{2005AJ....130.2598G}). This work uses data from the initial `$\alpha.40$' data release, consisting of 15,855 galaxies. The details of the survey selection can be found in \citealt{2011AJ....142..170H} -- in brief, however, the current data release of the survey covers 2800 deg$^{2}$ over two regions of the sky with a beam size of $3.5'$, and selects galaxies out to $\sim 250$ Mpc ($z \sim 0.05$). For such nearby galaxies, the SDSS fibre often only covers a fraction of the galaxy; see Appendix A for a discussion of potential aperture biases caused by the SDSS fibre. With a large number of sources and good sensitivity, the ALFALFA  $\alpha.40$ is an ideal resource for examining large scale trends in atomic gas content. 

This database of \HI-observed galaxies was then cross-matched with the SDSS spectroscopic survey \citep{2009ApJS..182..543A}. To cross-match the two databases, we defined a search box around each detected \HI\ position of 10 arc seconds in RA and DEC, and a redshift interval $\Delta z = 0.001$. We discarded any sources that returned either zero matches, or more than one unique match. The total number of galaxies in the ALFALFA-$\alpha.40$ survey appearing in the SDSS spectroscopic database is 8930. 

Of these galaxies, it was necessary that all lines required for metallicity diagnostics were detected. We discard any galaxies with a H$\alpha$ S/N lower than 25, which has the effect of ensuring that we have the full complement of lines required for our metallicity diagnostics. As discussed further in \S\ref{sec:met}, we use two metallicity diagnostics: the [N{\sc ii}]$\lambda 6584$/H$\alpha$ ratio, and the well-known `R23' parameter, defined as ${\rm R23} = ({\rm [OII]\lambda3727} + {\rm [OIII]\lambda4958,5007} )/{\rm H\beta}$.

Of the initial 8930 galaxies, 66\% have a H$\alpha$ S/N greater than 25, leaving us with 5952 galaxies. 

Following \cite{2010MNRAS.408.2115M}, we also discard galaxies for which our two metallicity diagnostics suggest metallicities differing by $>0.25$ dex, any galaxies which are `AGN dominated' using the BPT cut presented by \cite{2003MNRAS.346.1055K}.

After these cuts, the final \HI\ sample consists of 4253 galaxies. This sample is nearly 20 times larger than the sample used by Hughes et al. (2012) for a similar study, enabling us to achieve unprecedented statistical significance, and to investigate a range of galaxy properties unexplored by previous studies. 

For all galaxies appearing in the SDSS, total stellar masses (i.e., corrected for the effect of the fibre aperture) are taken from \cite{2003MNRAS.341...54K}. We have calculated star formation rates within the SDSS fibre, allowing direct comparison with Mannucci et al. (2010). Star formation rates are therefore calculated using the luminosity of the extinction corrected H$\alpha$ emission line, using the H$\alpha$/SFR conversion of \cite{1998ARA&A..36..189K}, modified to a \cite{2003PASP..115..763C} IMF using a conversion factor of 1.58 (this is the same star formation conversion adopted by Mannucci et al. 2010). 

\subsection{H$_2$ sample}
\label{sec:h2}

\begin{figure*}
\centering
\mbox
{
  \subfigure{\includegraphics[width=8.5cm]{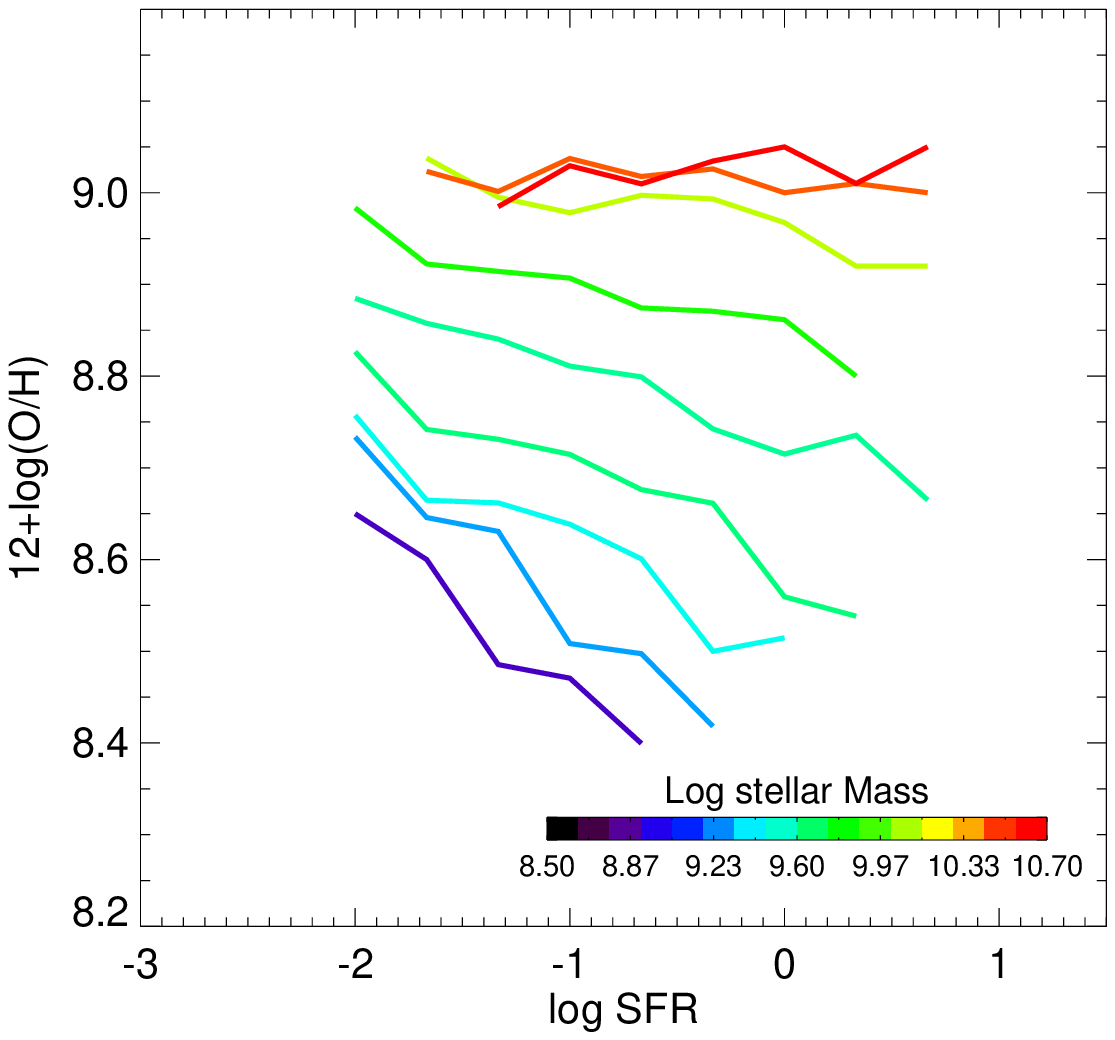}}
  \subfigure{\includegraphics[width=8.5cm]{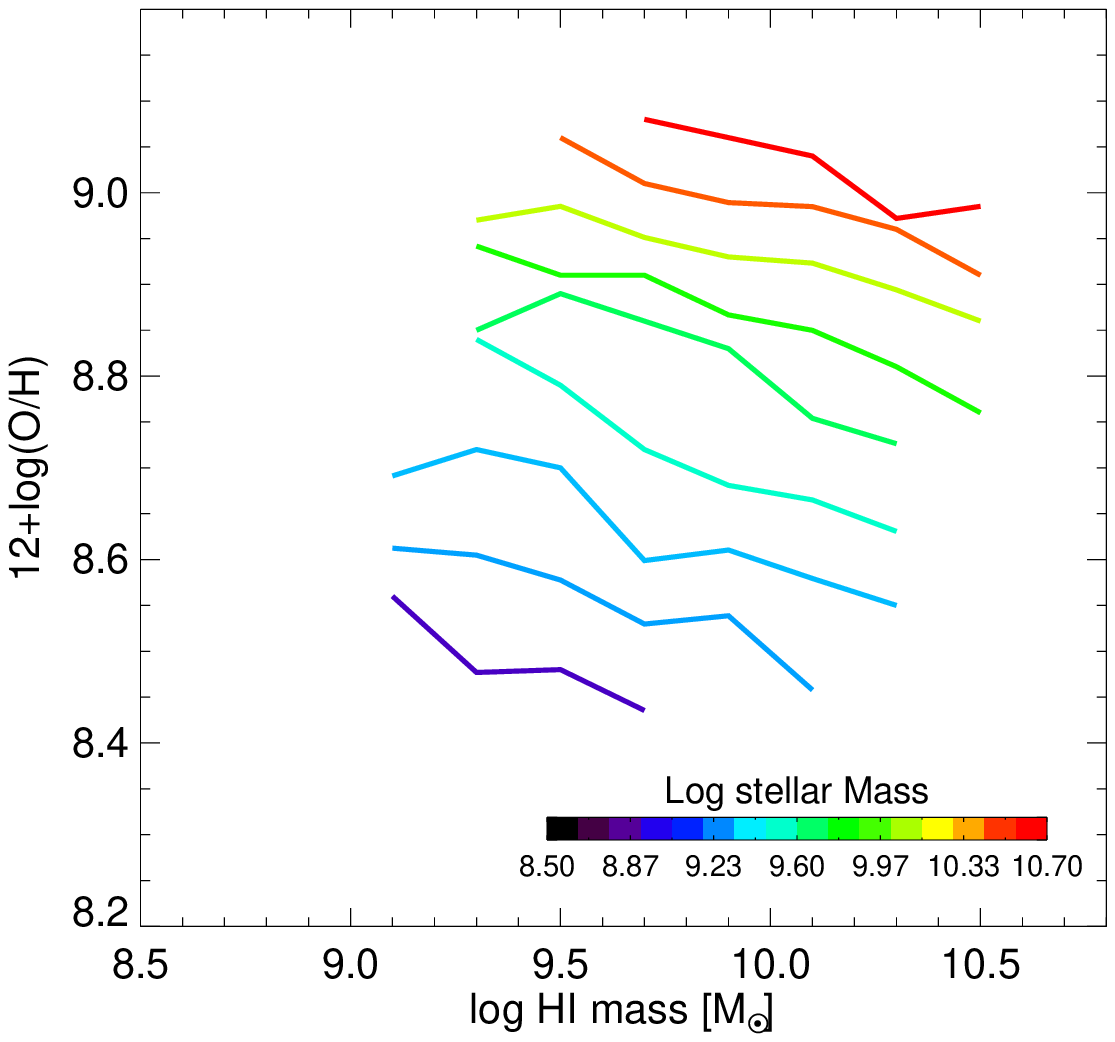}} 
}
\caption{The metallicity-SFR (left panel) and the metallicity-\HI\ (right panel) relation. The coloured lines show the mean trends for galaxies in bins of stellar mass. In each plot, a systematic dependence of metallicity on both SFR and \HI\ mass can be seen. At the highest stellar masses there is little dependence of metallicity on SFR, while the dependence of metallicity on \HI\ mass is seen across the full range of the sample.}
\label{fig:MZ-lines-alt}
\end{figure*}

The study of the molecular phase of the ISM is more observationally challenging than the atomic phase. Molecular hydrogen lacks a permanent electric dipole, necessitating the use of a proxy molecule (typically CO), from which the mass of H$_2$ can be extrapolated. The observation of CO is still tricky, however, and extragalactic CO surveys remain less populous than \HI\ surveys by well over an order of magnitude. The result, of course, is that samples dealing with molecular hydrogen observations are statistically impoverished relative to surveys of \HI. 

To minimise this issue, we assemble a sample of galaxies from the literature that have both compatible metallically diagnostics (i.e. R23 and [NII]/H$\alpha$) and CO data available. These additional galaxies have physical parameters (such as stellar mass) derived differently to the larger samples of SDSS galaxies.  Due to the varying selection parameters, we take care to separately identify all galaxies which are not part of the large surveys.

\subsubsection{COLDGASS sample}

The primary source of H$_2$-observed galaxies in this work is the COLDGASS survey \citep{2011MNRAS.415...32S}. COLDGASS is the H$_2$ subset of the GASS survey \citep{2010MNRAS.403..683C}, a study designed to obtain deep \HI\ data for a sample of relatively massive $z\sim0$ galaxies. The initial data release of COLDGASS \citep{2011MNRAS.415...32S} consists of 307  galaxies observed in CO, with a detection rate of 54\%. 

After applying the same selection criteria as applied to the ALFALFA sample above (i.e. a robust SDSS counterpart, well-detected metallicity indicators, lack of `AGN domination', and H$\alpha$ S/N $>25$, along with a CO detection), we are left with 72 galaxies. However, a further 33 of these galaxies are have highly inconsistent metallicity indicators; the [N{\sc ii}]$\lambda 6584$/H$\alpha$ ratio exceeds the `R23' parameter by $>0.5$ dex (in fact, in all of these cases the metallicity derived from the [N{\sc ii}]$\lambda 6584$/H$\alpha$ ratio simply saturates at the maximal value allowed by the calibrations). These galaxies most likely host LINERs or some weak Seyfert nucleus that contaminates the emission lines. In either of these cases the metallicity inferred from emission lines is not reliable. After excluding the COLDGASS galaxies with saturated [N{\sc ii}]$\lambda 6584$/H$\alpha$ metallicities, we are left with 39 COLDGASS galaxies observed in CO. 

Galaxies in the COLDGASS sample were cross-matched against the SDSS database as described above, which again provided star formation rates and stellar masses, just as for the ALFALFA sample. 

\begin{figure*}
\centering
\mbox
{
  \subfigure{\includegraphics[width=8.5cm]{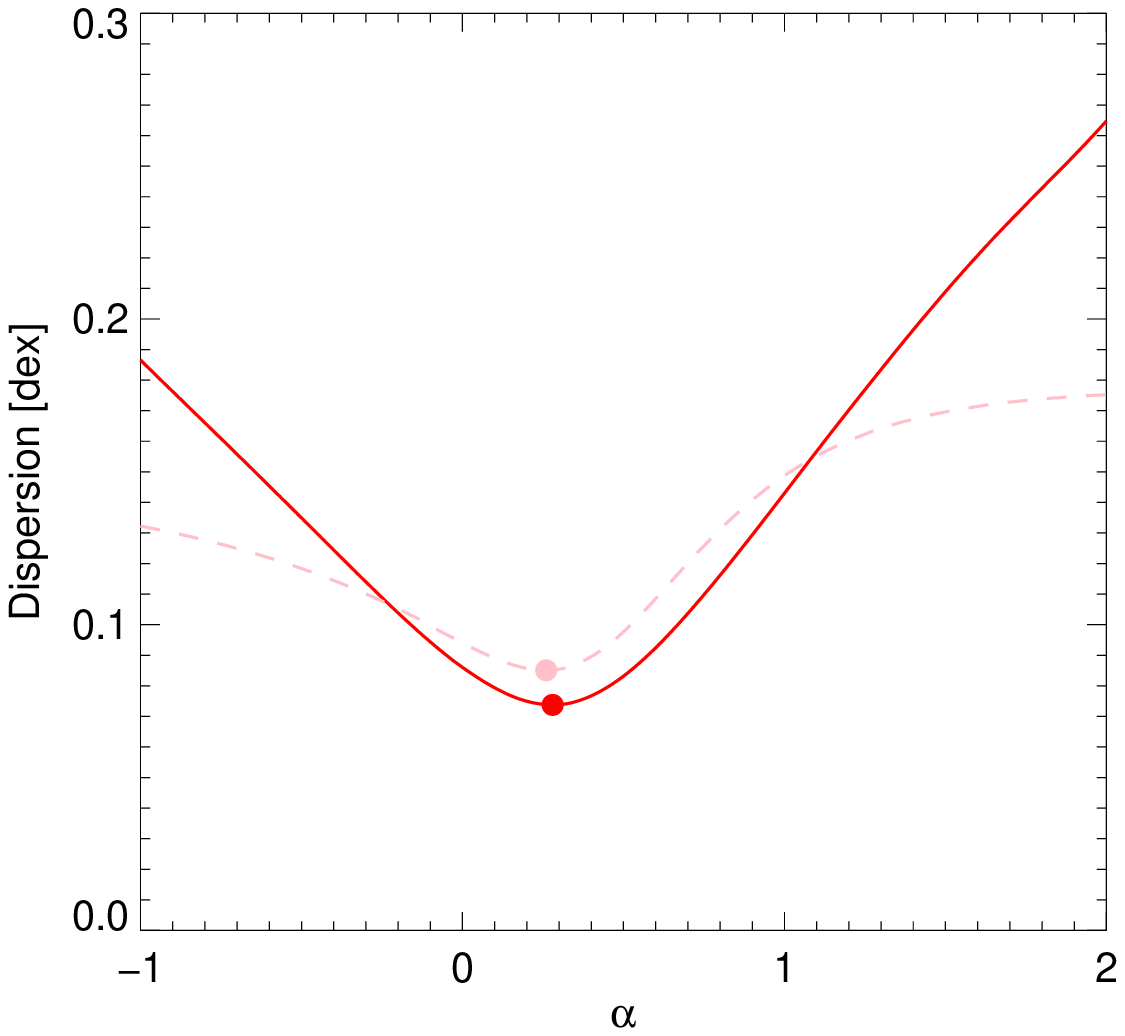}}
  \subfigure{\includegraphics[width=8.5cm]{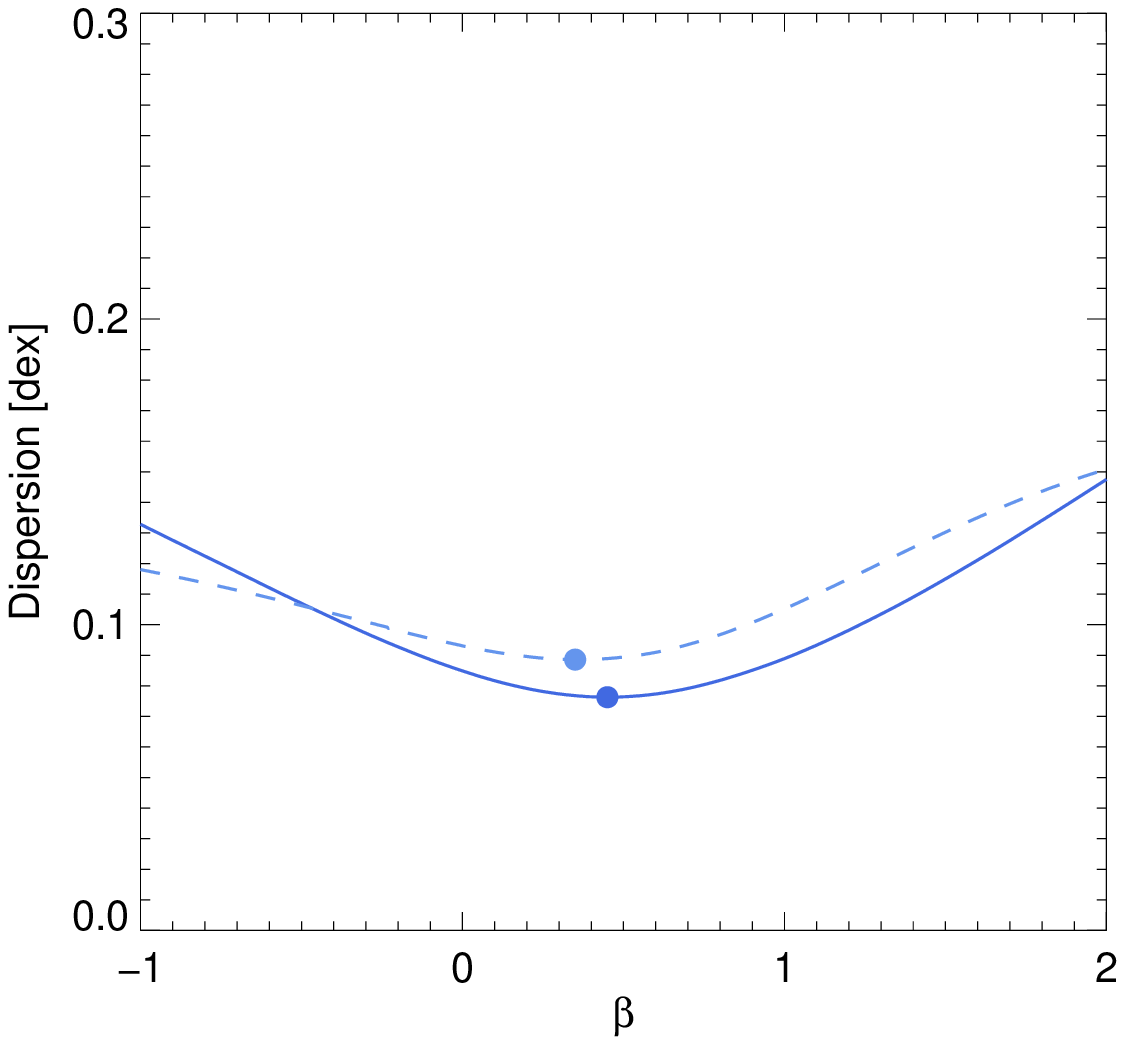}} 
}
\caption{The mean scatter in the `fundamental metallicity relation' as a function of the (SFR) parameter $\alpha$ (left panel, and Eq. 2) and the (\HI) parameter $\beta$ (right panel, and Eq. 3). Solid lines show the dispersion defined as the scatter around the polynomial best fit to the ($\alpha=0$) M$_*$-Z relation, taken from Mannucci et al. (2010). Dashed lines show the dispersion defined as the scatter around a best fitting fourth-order polynomial, recalculated at each $\alpha$. In each case, the dispersion in the mass-metallicity relation can be reduced by including an additional dependence on either SFR or \HI\ mass. The minimum dispersion reached for the SFR and \HI\ relations is 0.073 dex and 0.075 dex (respectively). We estimate that the uncertainly on the metallicity is $\sim 0.05$ dex (excluding systematics).}
\label{fig:alpha_scatter}
\end{figure*}

\subsubsection{Other CO galaxies}

As explained above, the number of galaxies currently observed in CO is approximately two orders of magnitude smaller than those observed in \HI. As a result, our H$_2$ sample is necessarily far smaller than our atomic gas sample. To go some way towards rectifying this, for the purposes of this study we also include a number of local galaxies that have been observed in CO, but are not part of a single homogeneous survey (like COLDGASS). Note that we take care to identify these galaxies separately in all plots hereafter, in order to avoid confusion.

We include an additional 23 literature galaxies observed in CO. These galaxies were primarily assembled from the literature by \cite{2012AJ....143..138S} and  \cite{2009MNRAS.394.1857O}. While \cite{2012AJ....143..138S} give metallicities taken from a variety of sources (primarily \citealt{2010ApJS..190..233M}), in order to remain as consistent as possible we re-derive metallicities using the method described above (i.e., taking the mean of the metallicity derived using the [N{\sc ii}]$\lambda 6584$/H$\alpha$ ratio and R23). 
Where the full complement of lines were unavailable, making this method impossible, we adopted the metallicity given by a single diagnostic -- either R23 or the [N{\sc ii}]$\lambda 6584$/H$\alpha$ ratio. This was only the case for 2 of these galaxies, however.

Most of the galaxies assembled by \cite{2012AJ....143..138S} appear in the SINGS survey \citep{2003PASP..115..928K}. They have stellar masses derived using SED fitting, given by \cite{2009A&A...507.1793N}\footnote{We have scaled the given stellar masses to a Chabrier IMF.}, and \HI\ masses from The \HI\ Nearby Galaxy Survey, THINGS \citep{2008AJ....136.2563W}. 

5 of our supplemental CO galaxies did not have SED-derived stellar masses available -- stellar masses for these 5 galaxies were derived using $K$-band magnitudes (as listed in the NASA ADS), convolved with a ($B - V$) colour-appropriate mass-to-light ratio, following the stellar models given by \cite{2001ApJ...550..212B}. In addition, 7 of our supplemental CO galaxies did not have a THINGS \HI\ mass listed -- for these galaxies, \HI\ masses were derived using the 21 cm magnitude taken from the online database HYPERLEDA.  

To summarise, the local molecular gas sample consists of 62 galaxies, 39 of which are taken from the homogenous, SDSS-crossmatched COLDGASS database, with physical parameters derived identically as for the ALFALFA-\HI\ sample described above. To improve our sample size, we have added 23 local galaxies taken from smaller CO surveys. These secondary galaxies have physical parameters derived using different methods to the SDSS samples, and therefore do not constitute a homogeneous dataset. 

The advantage of CO observations, in contrast with observations of atomic hydrogen, is that it can be detected out to cosmological distances with current instrumentation. The detection of \HI\ at redshifts beyond $z\sim 0.3$ will not be possible until the commissioning of the Square Kilometer Array (see \S\ref{sec:discussion} below). In order to examine the relationship between H$_2$ content and metallicity at high redshift, we also include a sample of high-$z$ star forming galaxies which have well characterised stellar masses, CO masses, and metallicities. We use a sample of `sub-millimeter galaxies' (SMGs), which have metallicities estimated using their [N{\sc ii}]$\lambda 6584$/H$\alpha$ ratio (using data from \citealt{Swinbank:2006aa}), molecular gas masses taken from \cite{2013MNRAS.tmp..563B}, and stellar masses derived by \cite{Hainline:2011lr}.

\subsubsection{H$_2$ gas masses}

The CO-to-H$_2$ conversion factor ($\alpha _{CO} = {\rm M}_{\rm H2}/{\rm L}'_{\rm CO(1-0)}$),
is one of the main sources of uncertainty in determining the
mass of molecular gas from the CO luminosity. The main issues involved in the determination
of $\alpha _{CO}$, and its dependence of the various galaxy parameters, are reviewed
in Bolatto et al. (2013). For our study, a central concern is the strong dependence of the
conversion factor on metallicity, which can obscure trends between H$_2$ mass and metallicity. We discuss this issue further in \S\ref{sec:h2fmr} below.
In this work, we have adopt a metallicity-dependent conversion factor, as given by Schruba et al. (2012):

\begin{equation}
\log \alpha_{\rm CO} = \log(A) + N(12 + \log{\rm O/H}) - 8.7,
\end{equation}

with $A=8.2 \pm 1.0$ and $N=-2.8 \pm 0.2$. This form for $\alpha_{\rm CO}$ is very similar to that given in other recent works (e.g. \citealt{2011MNRAS.418..664N}; \citealt{Genzel2011aa}), especially at the relatively high metallicities ($12 + \log{\rm O/H} > 8.5$) probed by our sample.

\subsection{Metallicity determination}
\label{sec:met}

The determination of the gas metallicity through strong line ratios has been widely
discussed in the literature. Some line ratios involving bright lines
such as [OIII]5007, [OII]3727, [NII]6584 and the Balmer lines, are sensitive to the metallicity,
although, in many cases with large scatter. The calibration of these `strong line ratios' has
been attempted both using empirical methods (e.g. by exploiting `direct' metallicity tracers)
and through photoionisation models, or through mixed methods (see discussion in
Kewley \& Ellison 2008 and in Maiolino et al. 2008). Each of these methods has advantages and
disadvantages, but it is beyond the scope of this paper to discuss the various metallicity calibrations in detail.
Since in this paper we wish to compare our results with the FMR inferred by Mannucci et al. (2010),
here we adopt the same calibration adopted by those authors: i.e. the strong line
calibrations given in Maiolino et al. (2008),
which use a combination of models and empirical calibrations.
In particular, we use the R$_{23}$ parameter and [NII]/H$\alpha$ ratio as metallicity diagnostics,
and we follow the same criteria as Mannucci et al. (2010), in that we take as the final metallicity the mean of the metallicities determined using the two methods.

\section{Results} 
\label{sec:results}

\subsection{The fundamental metallicity relation and HI content}
\label{sec:HIFMR}

Figure \ref{fig:MZ-lines} shows the mass metallicity relation for our sample of 4253 local galaxies appearing in both the ALFALFA $\alpha.40$ catalogue, and the SDSS. The grey shaded areas shows the region containing 64\% and 90\% (light grey and dark grey respectively) of all SDSS galaxies, as given by \cite{2004ApJ...613..898T}. 

The coloured lines in each panel show the mean metallicity of ALFALFA galaxies, separated into bins of SFR (left panel) and \HI\ mass (right hand panel). Bin sizes are 0.1 dex in \HI\, and 0.15 dex in SFR. Bin sizes were chosen to ensure sufficient numbers of galaxies in each bin; the number of galaxies per bin ranges from 21 (at the high end of SFR and \HI\ mass) to 841.

The left hand panel, displaying the mass-metallicity relation separated into bins of SFR, shows data which is functionally equivalent to the result presented by Mannucci et al. (2010) -- albeit with a smaller sample size. The global trend reported by those authors is clearly evident, with metallicity increasing with stellar mass, and decreasing with SFR at a constant stellar mass. 

In the right panel of Fig. \ref{fig:MZ-lines}, it can immediately be seen that there is also a strong, systematic influence of \HI\ mass on the mass metallicity relation, in much the same way as emerges with the SFR. At a given stellar mass, galaxies with greater \HI\ masses have proportionally lower metallicities. This effect can be seen down to the lowest stellar masses probed ($ \log ({\rm M}_{*} / {\rm M}_{\sun}) \sim 8.7)$ -- below which, the number of galaxies per bin becomes small, and any trends become unreliable). 

Interestingly, the two relations show markedly different behaviour at the highest stellar masses. As shown in Fig. 1, and as noted by Mannucci et al. (2010), the correlation between metallicity and SFR weakens towards the highest stellar masses, and the most massive galaxies in the sample ($\log \; ({\rm M}_{*} / {\rm M}_{\sun}) > 10.5$) show no correlation between metallicity and SFR. Conversely, the correlation between metallicity and \HI\ mass (as can be seen in the separation of the various \HI\ mass bins in metallicity space) is evident across the full range of stellar masses. 

This can be better seen in Fig. \ref{fig:MZ-lines-alt}, which shows metallicities plotted as a function of both SFR (left panel) and \HI\ mass (right panel). The mean metallicity is shown for bins of varying stellar mass. Less massive galaxies show a strong inverse dependence of metallicity on SFR, while the most massive galaxies -- $\log ({\rm M}_{*} / {\rm M}_{\sun}) > 10.5$ -- show an approximately constant metallicity with SFR. When considering the correlation between metallicity and \HI\ mass, however, it can be seen that the correlation persists even when considering the highest stellar mass bin. Even the most massive galaxies in our sample exhibit decreasing metallicity with increasing \HI\ mass. 

Another important point to take from Fig. \ref{fig:MZ-lines-alt} is the gradient of the metallicity-\HI\ mass slope. It can be seen that \HI\ mass correlates inversely with metallicity at all stellar masses, with a slope that is approximately constant with stellar mass. Importantly, this slope is significantly greater than $-1$ for all stellar mass bins. A slope of $-1$ might be expected if the sole physical process driving the FMR was ISM dilution by infalling metal-poor  gas (such that increasing the \HI\ mass by an order of magnitude decreases the value of $\log$(O/H) by a concomitant order of magnitude). Instead, the slope of the inverse metallicity-\HI\ mass correlation is closer to $-0.15$ (in the log-log space of Fig. 2), and is approximately constant across the stellar mass range probed by our sample.

The fact that the metallicity dependence on \HI\ mass is essentially the same at all stellar masses suggests that the same mechanisms responsible for metal enrichment and dilution are in place across galaxies of all masses.



\subsection{The dispersion in the HI-FMR}

\begin{figure}
\centering
  \includegraphics[width=8.5cm]{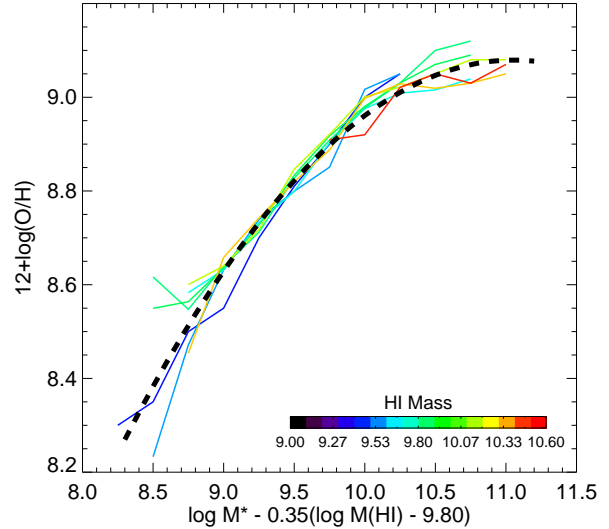}
\caption{The edge-on projection of the \HI-FMR, which minimises the scatter in metallicity. As in Fig. 1 (right), coloured lines show the mean trend for different \HI\ mass bins. The quadratic fit to the \HI-FMR (given in Eq. 4) has been overplotted as a black dotted line between $8.5 < \eta_{\beta} < 11$.}
\label{fig:HI_fmr}
\end{figure}

\begin{figure*}
\centering
\mbox
{
  \subfigure{\includegraphics[width=8.5cm]{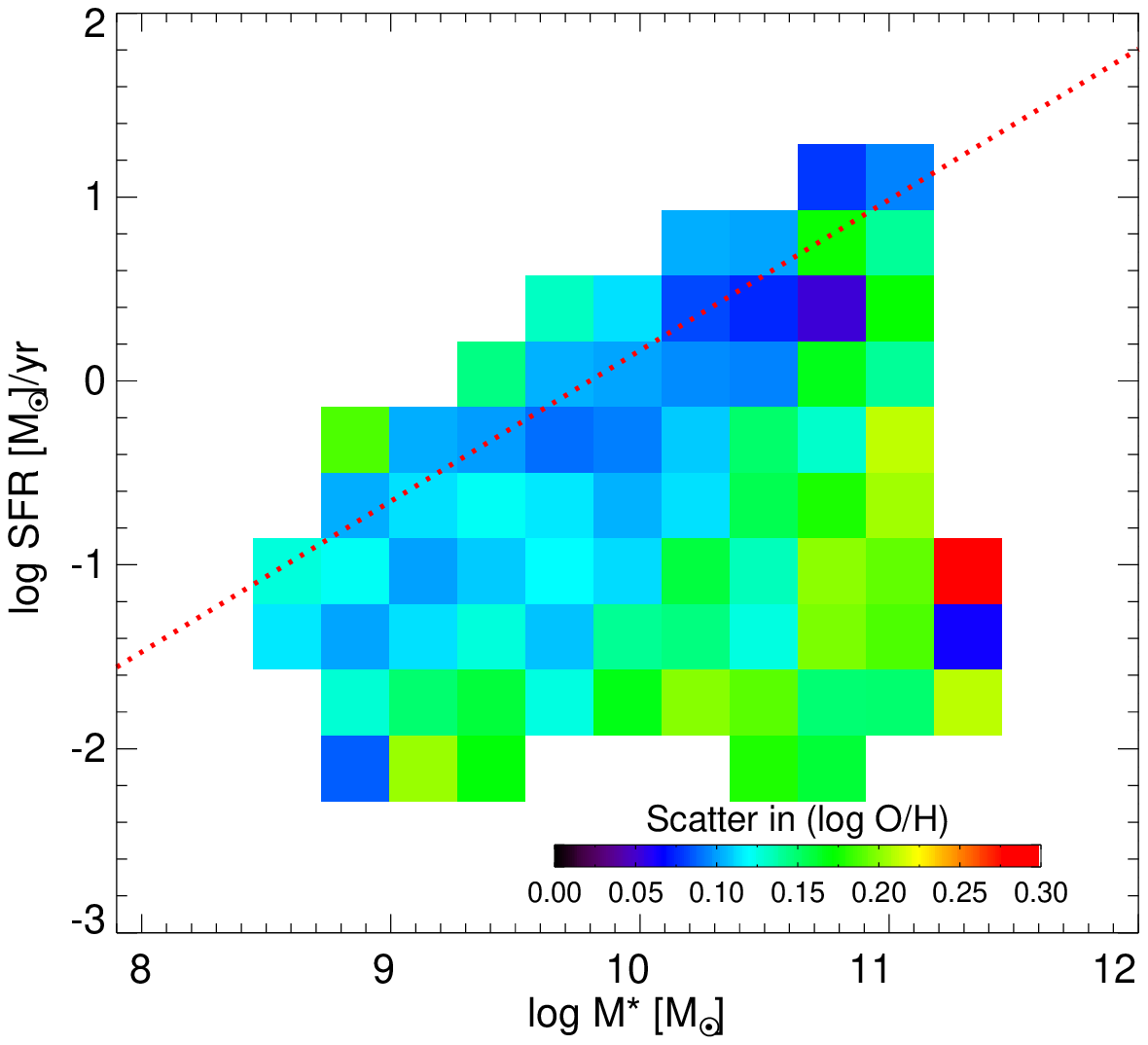}}
  \subfigure{\includegraphics[width=8.5cm]{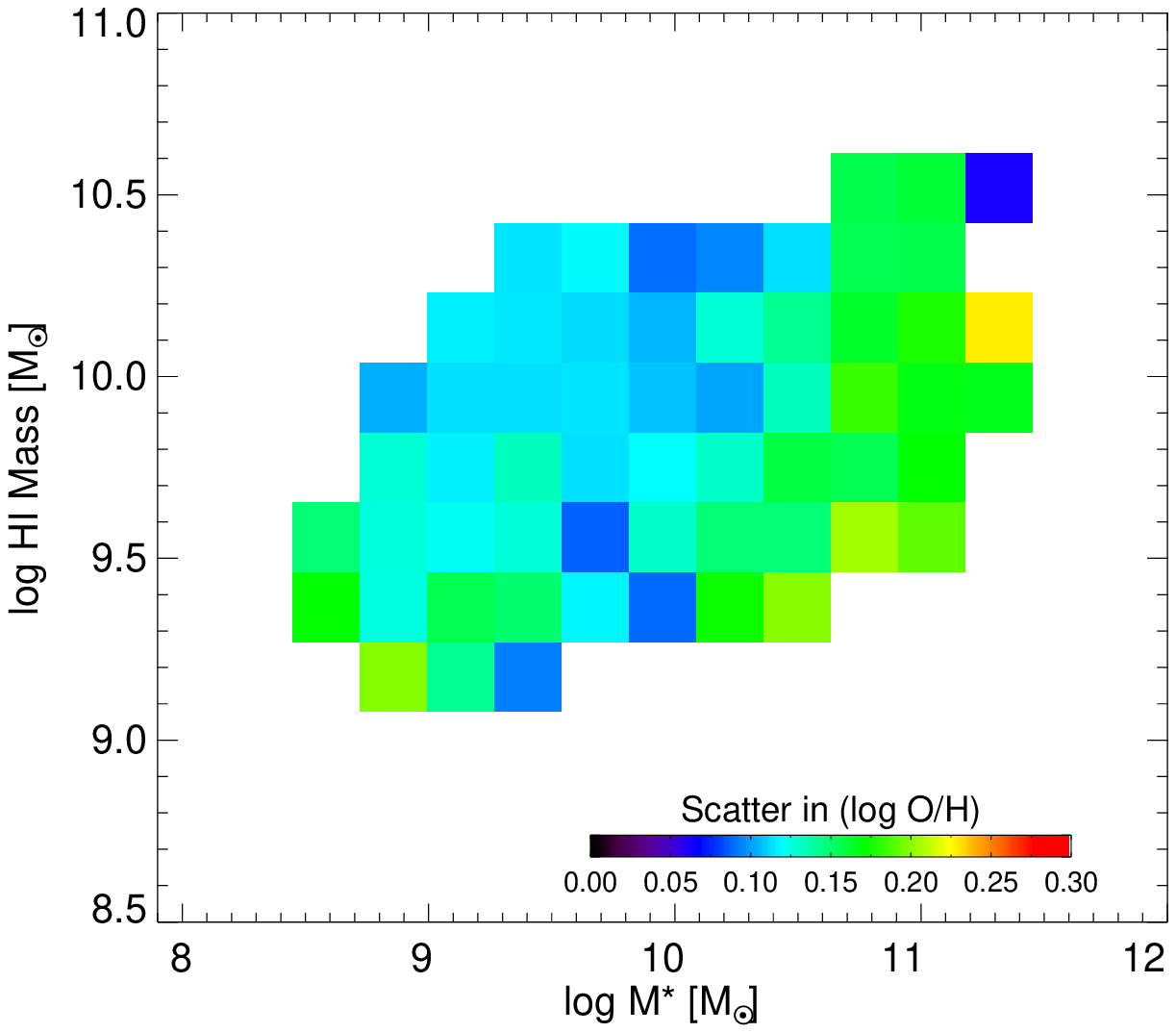}} 
}
\caption{Plots showing the dispersion in metallicity in the M*-M(\HI) plane ({\it left panel}) and the M*-\sfrA\ plane ({\it right panel}). The red dotted line in the right panel shows the SFR `Main Sequence', defined at $z=0$ by Elbaz et al. (2007).}
\label{MZ-disp-2d}
\end{figure*}

The recent papers that defined the FMR (\citealt{2010MNRAS.408.2115M}; \citealt{2010A&A...521L..53L}) defined the relation as a distribution of galaxies in a three dimensional parameter space defined by stellar mass, metallicity, and SFR. The three main `projections' of this fundamental parameter space, then, are the mass-metallicty relation, the mass-SFR relation, and the SFR-metallicity relation. \cite{2010MNRAS.408.2115M} consider a projection of this parameter space which minimises scatter and is not orthogonal to any axis (and therefore does not marginalise out one of the three parameters). 

They define a variable $\mu_{\alpha}$ which combines both stellar mass and SFR,  

\begin{equation}
\mu_{\alpha}= \log({\rm M})_* - \alpha \log ({\rm SFR}),
\end{equation}

and find that a value of $\alpha \sim 0.32$ minimises the scatter in the metallicity-$\mu$ plane.

We take a similar approach in finding the optimal projection of both the SFR-FMR and the \HI-FMR. For finding the optimal projection of the \HI-FMR, we adopt a parameter $\eta_{\beta}$ which combines both the stellar mass and \HI\ mass:

\begin{equation}
\eta_{\beta}= \log({\rm M})_* - \beta \log ({\rm M(\HI)} - 9.80).
\end{equation}

(The subtracted factor of 9.80 is equal to the (log) median \HI\ mass, and simply adds a linear scaling offset to $\eta$, to produce a range in $\eta$ that is comparable to the range in both $M_*$ and $\mu$.)

We analyse the scatter in the metallicity-$\eta_{\beta}$ relation as a function of varying $\beta$. Note that setting $\beta = 0$ causes $\eta$ to simply equal stellar mass (so that metallicity-$\eta_{\beta}$ becomes the standard mass-metallicity relation), and setting $\beta = 1$ causes $\eta$ to be the inverse atomic gas fraction $1 / f_{\rm HI}$ (if we define the atomic gas fraction $f_{\rm HI}$ as M$_{\rm HI}$/M$_*$). 

Figure \ref{fig:alpha_scatter} shows the mean dispersion in the metallicity-$\mu$ relation for SFR (left panel) and the metallicity-$\eta$ relation for \HI\ mass (right panel), as a function of varying the parameters $\alpha$ (left) and $\beta$ (right). In each, the point of minimum dispersion signifies the optimum projection of the FMR. We define dispersion in two slightly different ways. We have taken a fourth-order polynomial best fit to the standard $\alpha=0$ (and $\beta=0$) mass-metallicity relation following Mannucci et al. (2010): in each plot, the solid line shows the scatter around this original fit as $\alpha$ (and $\beta$) are varied. As a check, we also re-define a fourth-order polynomial fit to the $\mu$-metallicity (and $\eta$-metallicity) relation at each separate value of $\alpha$ (and $\beta$). The dashed line then shows the scatter around this fit, which varies as a function of $\alpha$ (and $\beta$). It can be seen from Figure \ref{fig:alpha_scatter} that the two definitions of scatter produce minima at essentially identical points.  

In the left hand panel, we show the data for varying $\alpha$, the projection of the SFR-FMR. We confirm that the result found by \cite{2010MNRAS.408.2115M} is also seen in our smaller sample -- that the optimal projection of the SFR-FMR is not with $\alpha=0$, but is at $\alpha = 0.28$.

We also demonstrate, for the first time, that the FMR also manifests equally when considering the influence of
\HI\ content on the mass-metallicity relation. The optimal projection of the \HI-FMR is found to be with $\beta =
0.35$. That is, the metallicity of galaxies in the local Universe can be well described
by the combination of the stellar and \HI\ masses, in the form of
$\eta_{0.35}= \log({\rm M})_* - 0.35 \log ({\rm M(\HI)} - 9.80)$. Fig.~\ref{fig:HI_fmr} shows the metallicity distribution of the
various bins in our sample as a function of this new parameter $\eta_{0.35}$, clearly showing that dispersion
is minimised along this projection and that
the metallicity depends solely on the parameter $\eta_{0.35}$.

We have fitted the overlapping curves of metallicity versus $\eta_{0.35}$ with the following
quadratic function:

\begin{equation}
12 + \log({\rm O/H} ) = -3.979(6) + 2.363(5)\eta_{0.35} - 0.1068(2)\eta_{0.35}^2 ,
\end{equation}

which traces the \HI-FMR closely between $8.5 < \eta_{\beta} < 11$. Numbers in parentheses indicate the uncertainty in the final digit of the fitted coefficients. 

It is also notable from Figure \ref{fig:alpha_scatter} that the dispersion in the FMR  varies more strongly with $\alpha$ (i.e. the SFR dependence) than of $\beta$ (the \HI\ mass dependence). That is, varying the projection angle of the FMR has more of an effect on the dispersion of the SFR-FMR than the \HI-FMR. Naively, this may suggest that the SFR-FMR is more tightly correlated than the \HI-FMR (consider the extreme situation where the dispersion is independent of projection, and the relation in Figure \ref{fig:alpha_scatter} is flat; this would imply no FMR exists, and the parameter has no influence on the mass-metallicity relation). However, the optimum projection of each relation produces a comparable minimum dispersion:  0.073 dex for the SFR, and 0.075 dex for the \HI, suggesting that the FMR can be expressed equally in terms of both SFR and \HI\ content. This dispersion is only slightly larger than the uncertainty on the metallicity, which we estimate to be $\sim 0.05$ dex (excluding systematic error). 


We suggest that this slightly stronger dependence on SFR could be due to the non-independence of the SFR and metallicity diagnostics; bright emission lines (particularly H$\alpha$) are instrumental in calculating both the metallicity and SFR of the SDSS galaxies in our sample, with the result that these two derived parameters are not truly independent. (This issue has been noted previously: see \citealt{2005ApJ...631..231P} and \citealt{2013ApJ...765..140A} for discussion). 

Indeed, Maiolino et al. (in prep.) have found that when the SFR is inferred from IR emission, the dispersion around the SFR-FMR becomes significantly larger. 
This result, combined with the results presented in this work, could suggest that the \HI-FMR is the more fundamental relation, with SFR-FMR being best understood as a by-product.

In addition, it must be noted that this direct comparison between the scatter in the SFR-FMR and the \HI-FMR is not entirely  fair, since in the former relation the spectroscopically-obtained parameters (metallicity and SFR) are measured within the same SDSS fibre aperture, while in the latter relation the \HI\ mass is measured via integrated 21cm emission --  not from the region sampled by the metallicity measurements. Compounding this effect is the fact that HI emission is often found out to large radii, beyond the stellar disk (see, e.g., \citealt{1994AJ....107.1003C}). As a result, it is likely that additional scatter is added to the \HI-FMR due to both the mismatched apertures and the extended \HI\ disk. Unfortunately, we cannot confirm this directly by extracting the \HI\ mass within the SDSS aperture. We can, however, attempt a comparison by checking the scatter in the SFR-FMR when using integrated SFRs (as inferred by Brinchmann et al. 2004), in order to introduce a similar `aperture mismatch' effect to that present in the \HI\ data. In this case, the dispersion around the SFR-FMR is higher ($\sim 0.1$ dex), and the relation between metallicity and SFR is weaker (with $\alpha$ closer to zero). It is likely, therefore, that if we were to use matched apertures, the lowest scatter would be evident when considering the \HI-FMR, further suggesting that the \HI-FMR is the more fundamental relation\footnote{See Appendix A for a discussion of fibre-based aperture effects relevant to SDSS metallicity measurements.}. 

Another way to visualise the scatter in these relationships (without needing to recourse to measuring the scatter around fitted functions) is to analyse the metallicity dispersion in the M$_*$-M(\HI) and the M$_*$-SFR planes. This is shown in Figure \ref{MZ-disp-2d}. The 1$\sigma$ scatter in metallicity is shown across planes of M$_*$-SFR (left panel) and M$_*$-M(\HI) (right panel), with bin sizes as before. The inset bar shows the scatter colour-coding up to a maximum value of 0.3 dex. In order to avoid misleading dispersions at the edges of the distributions caused by small number statistics, only bins containing $\geq 10$ galaxies have been plotted.

It is interesting to note that the metallicity dispersion is low (around or below 0.1 dex)
for intermediate stellar masses ($\rm 9<\log{(M_*/M_{\odot})}<11$) and \HI\ gas masses
($\rm \log{(M_{HI}/M_{\odot})}>9.5-10$), which may suggest that this region of the diagram
is predominantly populated by galaxies evolving smoothly, in equilibrium between SFR, gas inflow
and outflow. In contrast, massive galaxies, with $\rm \log{(M_*/M_{\odot})}\approx 11$ or higher
appear to have much higher metallicity dispersion, approaching 0.2 dex. Such a large dispersion
may reveal that massive galaxies are subject to a larger fraction of (minor) merging/interactions,
which make the metallicity deviate from the `regular' behaviour characterising secularly
evolving galaxies.

The region with low stellar and gas masses is also characterised by a rapidly increasing
dispersion. This property of low mass galaxies is likely a consequence of the fact that
the accretion of \HI\ clouds (responsible for the dilution of metals)
cannot any longer be considered as continuous
smooth events for low mass galaxies, since the mass of individual in-falling clouds is likely
comparable with the mass of gas already contained in the galaxy, hence producing a discontinuous
and scattered behaviour.



\begin{figure*}
\centering
\mbox
{
  \subfigure{\includegraphics[width=8.5cm]{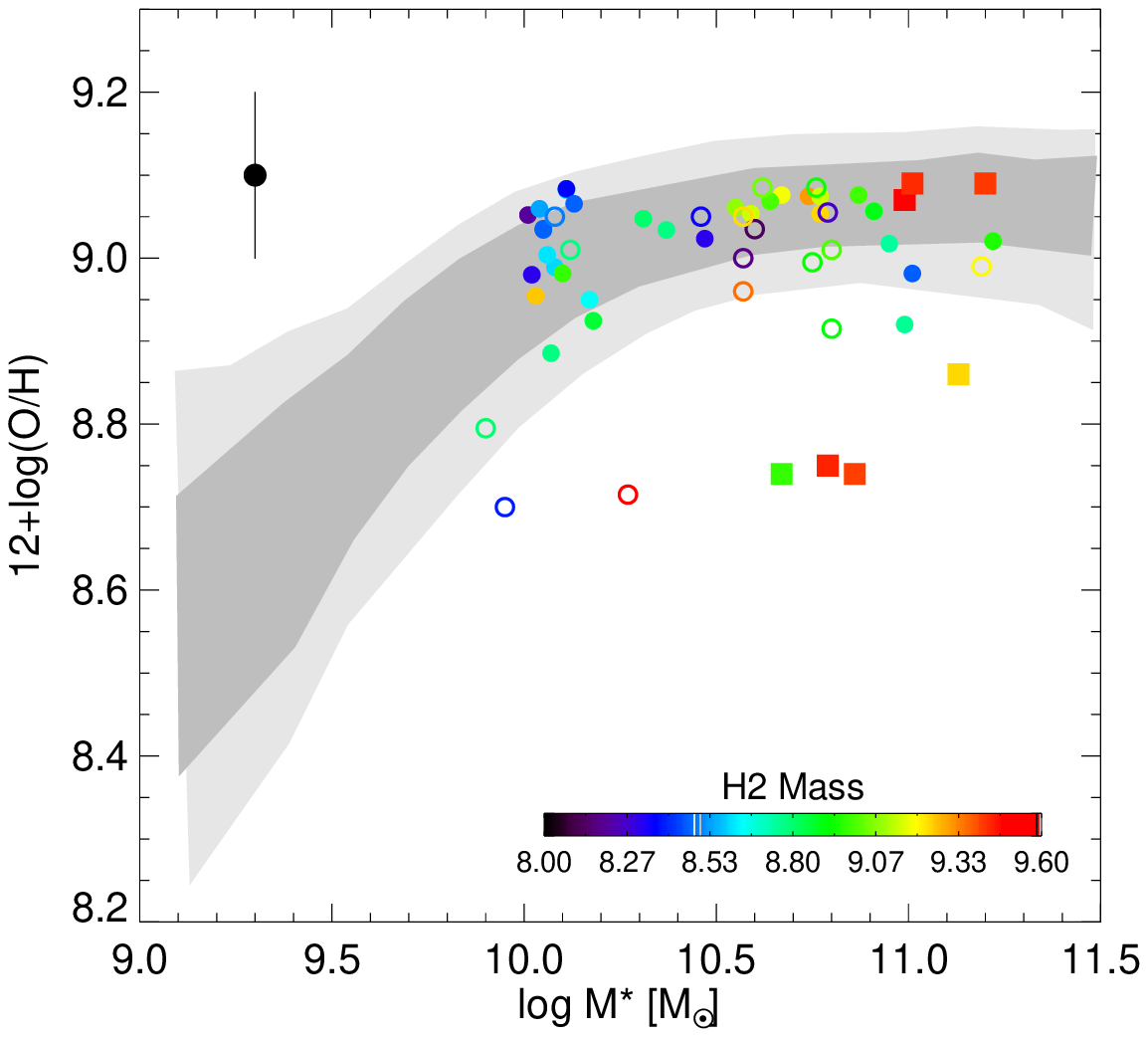}}
  \subfigure{\includegraphics[width=8.5cm]{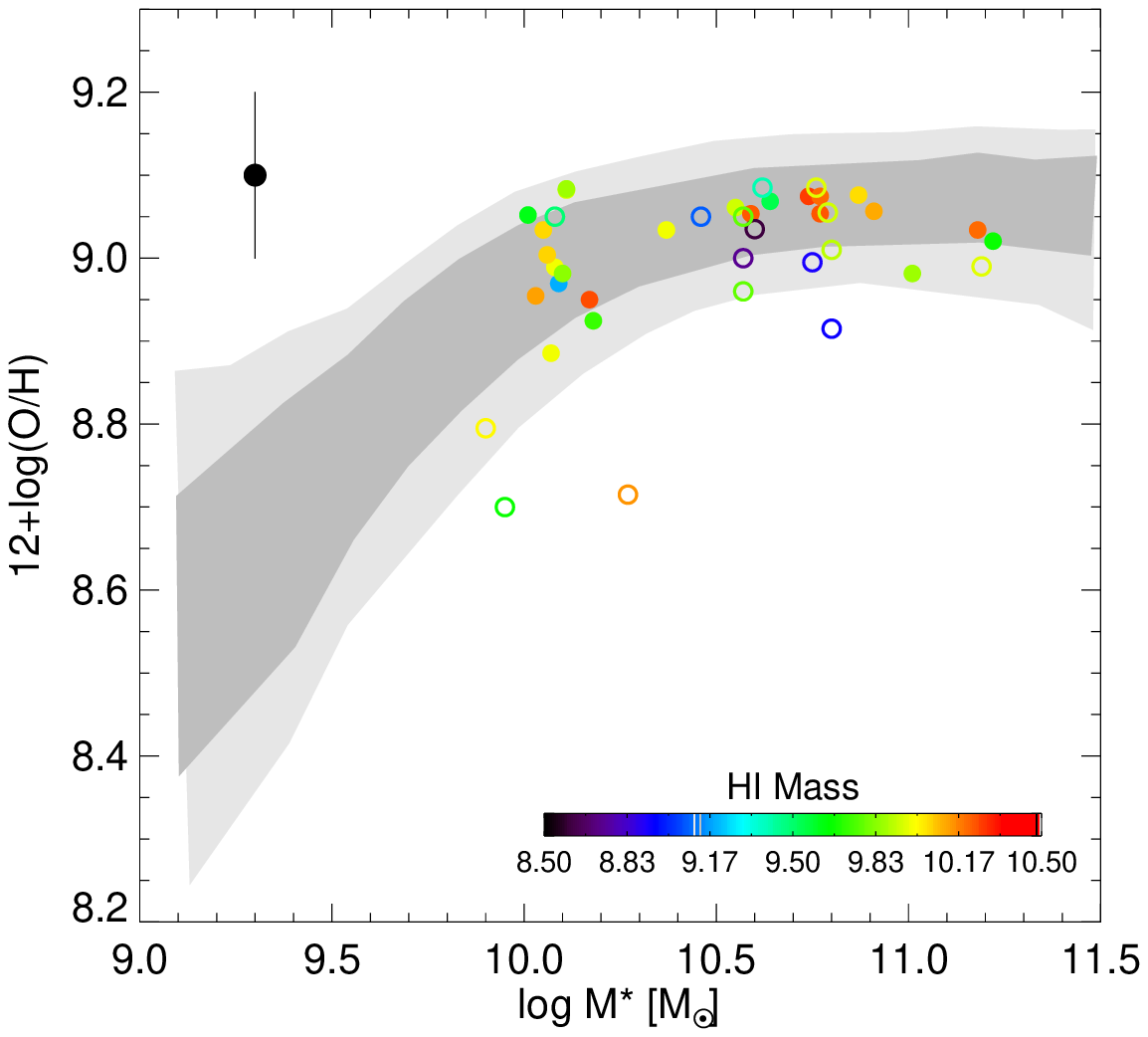}} 
%
}
\caption{The mass-metallicity relation for galaxies with CO observations. In each panel, galaxies appearing in COLDGASS (and therefore the SDSS) are plotted with filled circles. Galaxies gathered from the literature are plotted with open circles. High-$z$ SMGs are shown as squares. {\it Left panel}: galaxies colour-coded with their H$_2$ mass. A metallicity-dependent CO/H$_2$ conversation factor has been assumed (see text for discussion). {\it Right panel}: galaxies colour-coded with their \HI\ mass. In each panel, the black point in the top left shows the typical uncertainty in metallicity.}
\label{fig:h2}
\end{figure*}

\subsection{The fundamental metallicity relation and H$_2$ content}
\label{sec:h2fmr}

Figure \ref{fig:h2} shows the mass-metallicity relation for the 62 local galaxies in our sample that have been observed in CO. These include galaxies in the COLDGASS sample, which appear in the SDSS (and therefore have metallicities, stellar masses, and star formation rates calculated identically as for the ALFALFA sample above), as well as other galaxies from the literature. These two samples are differentiated by plotting with full and empty symbols (respectively). In addition, 9 high-$z$ SMGs with stellar masses, CO observations, and [N{\sc ii}]$\lambda 6584$/H$\alpha$ metallicities are also plotted as square symbols. Both panels show the mass-metallicity relation, with different colour-coding; the left panel shows the mass-metallicity relation colour-coded with H$_2$ mass, while the right panel colour-codes with \HI\ mass.

It can be seen that there is no apparent trend in metallicity between molecular gas rich and gas poor galaxies. Unfortunately, due to the small sample size, separating the galaxies into several bins of H$_2$ mass would not produce physically meaningful results. Instead, we can get a crude estimate of any possible trend by separating the sample into two broad categories: `molecular gas rich' galaxies (with H$_2$ masses greater than the median value ($\log {\rm M}_{\rm H2} = 9.35$), and `molecular gas poor' galaxies with H$_2$ masses lower than this. The mean value of the metallicity is the same for each subsample; $12 + \log ({\rm O/H}) = (9.00 \pm 0.09)$ for the H$_2$ rich subset, and ($9.00 \pm 0.09$) for the H$_2$ poor subset\footnote{Here, and throughout this discussion, errors represent a $1\sigma$ standard deviation in the relevant population, rather than an uncertainty in the quoted mean.}. The mean stellar masses for the two subsamples are similar, with mean stellar masses of log M(*) = $(10.53 \pm  0.38)$ and  $(10.37 \pm  0.38)$ for the H$_2$ rich and H$_2$ poor subsets, respectively. 

This result for molecular gas content stands in contrast to what we have seen above (a strong dependence of metallicity on \HI\ mass), and initially could suggest a lack of dependence of the metallicity on the molecular gas content. This would be a very surprising result, given that it is the molecular gas phase that is most intimately connected with star formation. However a similar lack of dependence is also seen in the \HI\ data for the H$_2$ sample. Again adopting the dual `gas rich/gas poor' binning, galaxies in our H$_2$ sample which are \HI\ rich have mean metallicities of $9.00 \pm 0.09$, whereas galaxies in the same sample which are \HI\ deficient have a mean metallicity of $8.99 \pm 0.09$. The same goes for the relationship with the total gas mass\footnote{M(gas) = 1.36(M(\HI) + M(H$_2$)), where 1.36 is a correction applied to account for interstellar helium.} ($9.01 \pm 0.08$ and $8.98 \pm 0.10$ for gas rich and gas poor galaxies, respectively). 

The lack of dependence of \HI\ content on the metallicity of our smaller (H$_2$-observed) sample, compared to the significant dependence observed in our large ALFALFA-SDSS sample (see \S\ref{sec:HIFMR} above), suggests that some combination of sample size and selection effects are obscuring any potential trend. 

The COLDGASS sample (which is a random selection drawn from the parent sample, GASS) includes only galaxies with M$* > 10^{10}$ M$_{\sun}$, and is selected to have a flat distribution in $\log {\rm M}_*$. This stellar mass coincides with the turnover in the mass-metallicity relation, with the result that the metallicity range of our H$_2$ sample is restricted relative to our (larger) \HI\ sample, providing less dynamic range over which any potential trends can be examined.  Furthermore, the 62 CO-observed galaxies remaining after our sample selection cuts (primarily the removal of AGNs and CO non-detections)  exhibit a particularly restricted metallicity range, with a standard deviation in 12+log(O/H) of $<0.1$ dex.

It is unlikely that CO non-detections are significantly affecting any of the results presented here. While 35\% (83/233) of the COLDGASS parent sample are listed as CO non-detections,  96\% of those CO non-detects (80/83) are removed from our sample by the selection cuts described in \S2.1 and \S2.2. Just three COLDGASS galaxies undetected in CO satisfy our selection criteria, and those three galaxies have metallicities consistent with the COLDGASS average. (The three CO non-detects do have a lower SFR, however, of $\log ({\rm SFR}) = -1.46 \pm 0.83$, compared to the detected sample average of $\log ({\rm SFR}) = -0.76 \pm 0.62$.)

A further complication in the attempt to test the existence of a molecular gas FMR is the strong dependence on the CO/H$_2$ conversion factor on the gas-phase metallicity. Although there have been several attempts to model and to observationally constrain the relation between the CO/H$_2$ conversion factor and gas-phase metallicity (e.g. \citealt{Genzel2011aa}; \citealt{2011MNRAS.418..664N}; Bolatto et al. 2013), such a dependence is so strong that uncertainties in the metallicity and/or deviations from the nominal relation adopted in this work, can strongly affect or wash out any real trend between molecular gas content, metallicity and stellar mass.

\section{Discussion}
\label{sec:discussion}

Several models have attempted to reproduce the Fundamental Metallicity Relation between
SFR, metallicity and stellar mass (e.g. \citealt{2012MNRAS.421...98D}; \citealt{2012arXiv1202.4770D}; \citealt{2011MNRAS.417.1013C}),
in terms of inflows, outflows and star formation efficiency. It is an important goal to
investigate whether the same models are able to reproduce the Fundamental Metallicity Relation
between atomic gas mass, metallicity and stellar mass. An attempt to model the metallicity
dependence on the \HI\ content
was already reported in \cite{2013arXiv1302.3631D}. Here we have
 provided a much more detailed and quantitative fundamental relation, which can be used
to test galaxy evolutionary models in detail. In particular, it is important
to verify whether current models can explain the metallicity dependence on \HI\ mass even
in massive galaxies (in contrast to what is observed for the SFR-metallicity relation)
and whether they can explain the large scatter observed in massive and in low mass galaxies.

It is beyond the scope of this paper to develop detailed models to interpret the \HI-FMR
and its scatter. We only note that the observed relation cannot be explained in terms
of a simple closed-box evolution (as suggested by other authors, e.g. \citealt{2012MNRAS.427.1075M}) where
the anti-correlation between gas mass and metallicity is a simple consequence of the ageing of the
galaxy (in which the gas is gradually consumed and enriched by star formation.) 

\begin{figure}
\centering
  \includegraphics[width=8.5cm]{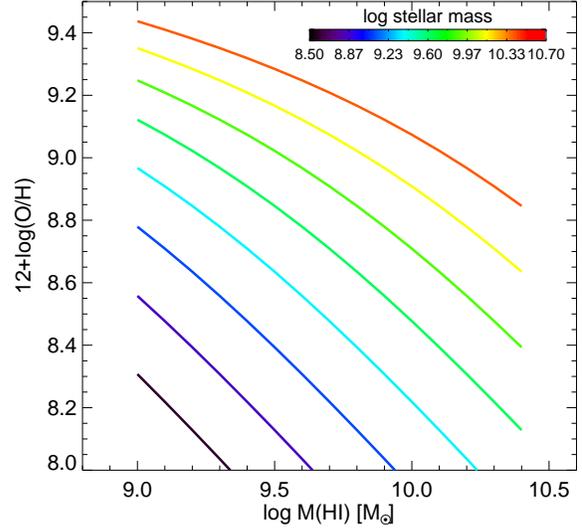}
\caption{The metallicity-\HI\ mass relation, shown for a range of stellar masses, derived using the `closed box' model given by Erb (2008). Compare to Fig. 2 (right panel). While the model reproduces some of the trends seen in the data (the inverse correlation between \HI\ mass and metallicity), the slope at a given stellar mass is far steeper in the models than is seen in the data. }
\label{fig:model}
\end{figure}

Fig. \ref{fig:model} shows the $\rm M_{HI}$ vs metallicity relation for different stellar masses, derived using a simple `closed box' model presented by Erb (2008)\footnote{See Eq. 11 in that work.}. For this toy model, we have assumed that all gas is atomic (that is, the gas mass is equivalent to the \HI\ mass), the fraction of mass remaining locked in stars, $\alpha = 1$, and the yield $y = 0.019$. It must be noted that varying these simple assumptions does not change the form of the results in any significant way. We have also converted between `metal content' $Z$ (defined as the mass fraction of elements heavier than helium) and 12+log(O/H):

\begin{equation}
12 + \log{\rm (O/H)} = \log \left( \frac{Z}{0.0126} \right) + 8.69
\end{equation}

The model data reproduce the basic form of the metallicity-\HI\ mass relation shown in Fig. 2 -- that metallicity is inversely correlated with \HI\ mass at a given stellar mass (and directly correlated with stellar mass at a given \HI\ mass). However, there are some significant differences which suggest that a simplistic `closed box' model is insufficient to explain the metallicity-\HI\ mass relation as revealed by the ALFALFA sample. The model produces a metallicity-\HI\ mass correlation with a gradient that is (a) significantly steeper than is seen in the data, and (b) dependent on stellar mass. The gradient of the model metallicity-\HI\  correlation varies between $-0.9$ for the lowest stellar mass galaxies, to $-0.4$ for the most massive. This stands in contrast to the trend seen in the ALFALFA data, where the gradient is approximately constant at $-0.15$ across all stellar mass bins. A `pure dilution' scenario, in which metal abundance is diluted by an infalling (pristine) gas cloud also fails to account for the data -- this scenario would result in a trend between \HI\ mass and metallicity even steeper than produced by the closed box model. 

It is clear from these model results that the most basic features of the data (the inverse correlation between \HI\ content and metallicity) can be reproduced by a simple `closed box' model, or dilution model. However, it is also clear that more complex modelling (likely involving both inflow/outflow of gas, and metal enrichment by star formation) is required to fully account for the flatter trends revealed in Fig. 2. It is possible that outflows are responsible for the dependence on mass, with infalls controlling the dependence with SFR; outflows are the dominant force in the chemical evolution of low mass galaxies, which (in the simple model) show the greatest discrepancy with the results presented here. 

In addition, further work is required to understand the role of molecular gas in these trends. While the CO data presented in this work do not allow us to draw conclusions about any potential correlation between molecular mass, stellar mass, and metallicity, due to the paucity of data we certainly have not been able to rule any such correlation out. Indeed, as star formation is more closely connected with molecular gas than atomic gas, the existence of the SFR-FMR would at least suggest the existence of a `molecular gas FMR'. Larger datasets, and a better understanding of the metallicity-dependence of the CO/H$_2$ conversion factor, will be required in order to explore this.

Finally, it is an important future goal to investigate whether the \HI\ FMR discovered here
also holds at high redshift. While hundreds of high-$z$ galaxies have had their gas-phase metallicities studied, 
detecting \HI\ at cosmological redshifts will only be feasible
with the SKA pathfinders (MeerKAT and ASKAP), and studying significant numbers of galaxies in \HI, spanning
a wide range of masses, will require the full deployment of the SKA.
Potentially, the ability to investigate the relationship between metallicity and molecular gas
will be within reach on a shorter timescale. 
Both in the local universe and at high
redshift larger samples of galaxies are expected to become available on short timescale thanks to ALMA,
and to upcoming extensive surveys using IRAM. However, the strong dependence of the CO-to-H$_2$ conversion
factor on metallicity, and the associated uncertainties, will likely remain a major issue
which may hamper our ability to identify and characterise a putative H$_2$ FMR.

\section{Conclusions}
In this work, we have examined the interconnection between the stellar mass, gas-phase metallicity, and gas content of local galaxies. Our main conclusions are as follows:

\begin{itemize}
\item{We have presented evidence that the mass-metallicity relation exhibits a strong secondary dependence on \HI\ mass, with \HI-rich galaxies being more metal poor at a given stellar mass. This trend is evident across the stellar mass range probed by our sample.}\\

\item{We introduce the \HI\ fundamental metallicity relation, an optimum projection of metallicity, stellar mass, and \HI\ mass, which reduces dispersion relative to the mass-metallicity relation. We expect that this is a fundamental relation, which lies behind the well-known SFR-FMR.}\\

\item{We also note that the dispersion in metallicity is lowest $<0.1$ dex for galaxies of intermediate stellar and \HI\ mass. We argue that this is the regime populated by secularly evolving galaxies in equilibrium between SFR, inflow, and outflow.}\\

\item{We assemble a smaller sample of galaxies observed in CO, but see no correlation between molecular gas content and metallicity. We argue that this does not rule out a H$_2$-FMR, but is due to a combination of selection effects and complications involving a metallicity-dependent CO/H$_2$ conversion factor.}\\

\item{Finally, we note that a simple `closed box' model is not sufficient to explain the trends in the \HI-FMR. Developing more sophisticated models to fully account for the behaviour between \HI\ mass, metallicity, and stellar mass is an important future goal.}

\end{itemize}

\section*{Acknowledgments}

We thank the referee for providing useful comments which have improved this manuscript. This research has made use of NASA's Astrophysics Data System. AM and FM acknowledge support from grant PRIN-MIUR 2010-2011 ``The dark Universe and the cosmic evolution of baryons: from current surveys to Euclid''. 

\bibliography{/Users/Matt/Documents/mybib}{}
\bibliographystyle{mn2e}

\appendix

\section{Aperture effects: the size of the SDSS fibre}
One potential source of bias that must be considered is the aperture effect of the SDSS fibre. Each SDSS fibre has a typical diameter of $\sim 3''$. At a distance of 100 Mpc (approximately the lower limit of our \HI\ sample), this corresponds to a physical extent of a few kpc. Clearly, such small apertures will only capture a small fraction of the total light from more extended galaxies, and aperture corrections must be applied to photometry in order to estimate integrated magnitudes. 

The existence of abundance gradients within individual galaxies could potentially cause metallicities calculated from small apertures to be biased. The galactic bulge typically exhibits a higher metallicity than the disc (see, e.g., \citealt{1999PASP..111..919H}), so measuring a metallicity using spectra taken from a fixed central aperture can cause abundance estimates to be biased high. 

\begin{figure}
\centering
  \includegraphics[clip=true, trim = 0cm 0cm 0cm 0cm, width=8.5cm]{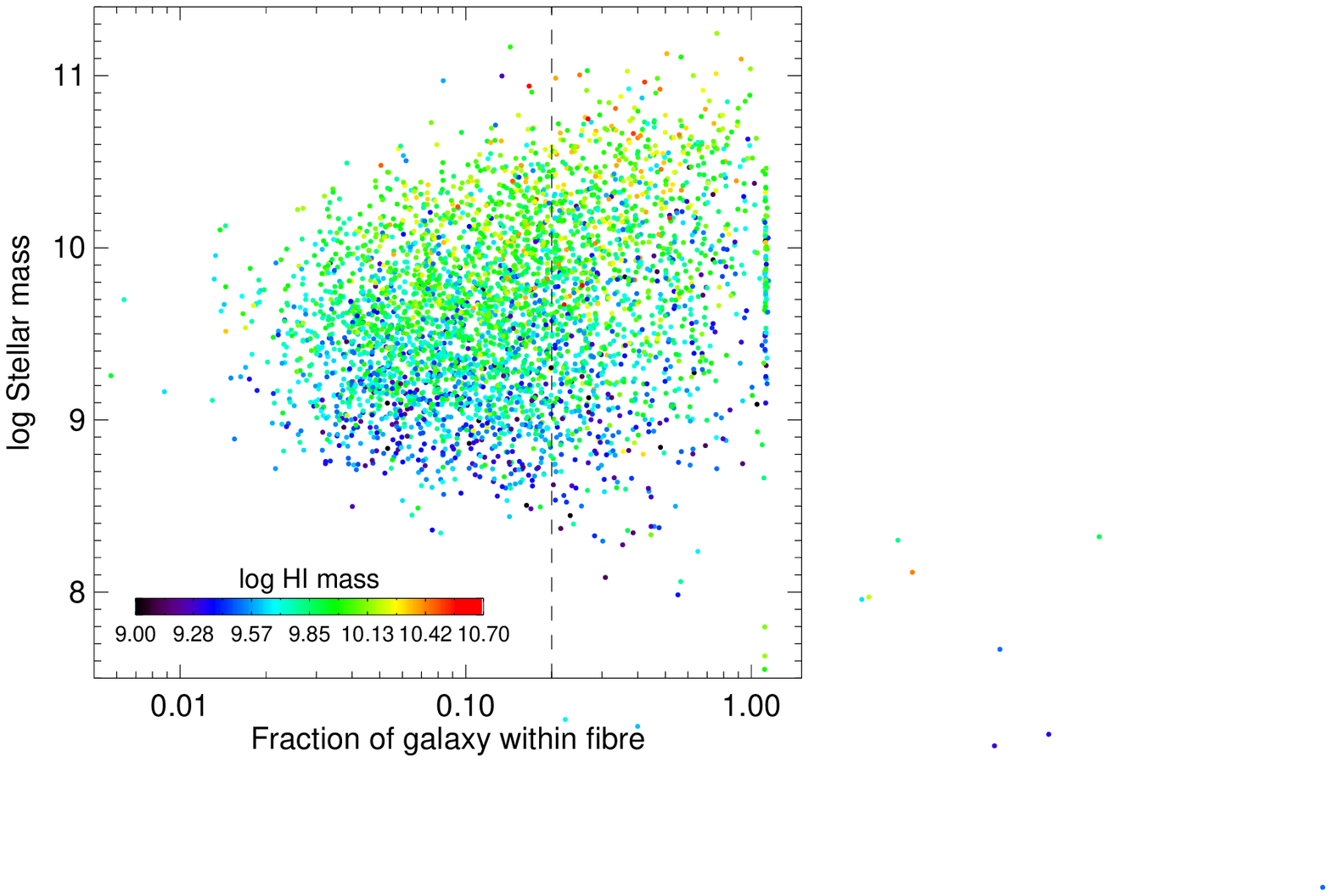}
\caption{The fraction of each \HI\ galaxy contained within the SDSS fibre, plotted against stellar mass. The fraction within the fibre was calculated by taking the ratio of the fibre-SFR to the total, aperture-corrected SFR. Kewley et al. (2005) estimate that a covering fraction of $>20\%$ (indicated with a vertical dashed line) is needed to avoid bias resulting from metallicity gradients.}
\label{fig:aperture}
\end{figure}

\cite{2005PASP..117..227K} estimate that a fibre covering fraction of $20\%$ is a useful benchmark, above which  abundances derived from a central aperture agree well with integrated values. We estimate the fraction of a galaxy covered by the SDSS fibre by taking the ratio of the fibre SFR and the total (i.e. aperture-corrected) SFR, as given by \cite{2004MNRAS.351.1151B} and the MPA-JHU group. Figure \ref{fig:aperture} shows the fraction of light within the fibre plotted against stellar mass for our ALFALFA-\HI\ sample. Approximately 1/3 (1413/4253) of the galaxies in our \HI\ sample have $>20\%$ of their light within the SDSS fibre, and should therefore be relatively unaffected by bias introduced by metallicity gradients. 

We have repeated the analyses in this work for this `aperture unbiased' subsample. While we lose some dynamic range in \HI\ mass by restricting our sample in this way, all our main conclusions (the existence of a \HI-FMR, with \HI-rich galaxies being more metal poor at all stellar masses) remain robust when considering only galaxies for which bias due to metallicity gradients is likely to be negligible.

\end{document}